\documentclass[nofootinbib,reprint,showkeys,showpacs,amsmath,amssymb,aps]{revtex4-1}
\usepackage{graphicx}
\usepackage[utf8,latin1]{inputenc}
\usepackage{dcolumn}
\usepackage{xcolor}
\usepackage{bm}
\usepackage{hyperref}
\usepackage{amsmath}
\usepackage{hyperref}
\usepackage{colortbl}

\allowdisplaybreaks[1]
\begin{document}


\title{DPSV trick for spherically symmetric backgrounds}

\author{S. Mironov}
\email{sa.mironov\_1@physics.msu.ru}
\affiliation{Institute for nuclear research RAS, Moscow, 117312, Russia}
\affiliation{Institute for Theoretical and Mathematical Physics, MSU, Moscow, 119991, Russia}
\affiliation{Kurchatov Institute, Moscow, 123182, Russia}

\author{V. Volkova}
\email{volkova.viktoriya@physics.msu.ru}
\affiliation{Institute for nuclear research RAS, Moscow, 117312, Russia}
\affiliation{Department of Particle Physics and Cosmology, Physics Faculty, MSU, 119991 Moscow, Russia}

\begin{abstract}
We discuss the approach suggested by Deffayet et al. (DPSV) to analysing the linearized perturbations in Horndeski theory in the case of a static, spherically symmetric background. In $\mathcal{L}_3$ subclass 
of Horndeski theories we prove the validity of the DPSV approach by showing that the original method corresponds to a specific gauge choice in the quadratic action for perturbations. We also show that in the case of a spherically symmetric background the DPSV trick does not work in a more general $\mathcal{L}_4$ Horndeski theory.

\end{abstract}

\maketitle

\section{Introduction}

In recent years healthy scalar-tensor theories with the second derivative terms in the Lagrangian, i.e. DHOST theories~\cite{Langlois1,Langlois2}, were proved to be a rich framework for all sorts of both cosmological scenarios 
and 
static, spherically symmetric backgrounds describing compact gravitational objects.  
In particular, DHOST theories and their subclasses, e.g. (beyond) Horndeski theories~\cite{Horndeski,GLPV}, are renown for admitting healthy Null Energy Condition (NEC) violation~\cite{RubakovNEC}, which makes them suitable for modelling cosmological scenarios without initial singularity (e.g. bouncing Universe or Genesis scenario) as well as non-singular gravitational objects like wormholes.

Today it is known, however, that 
these non-singular solutions within DHOST theories are generally exposed to pathological instabilities at the linearized level of perturbations unless a certain set of stability constraints is satisfied~\cite{nogoLMR,nogoKoba,nogoEM}. The latter means that a thorough stability analysis is usually required, based on studying either the equations for perturbations or the corresponding perturbed quadratic action. 

The most straightforward way to study stability at the linearized level is to plug the chosen parametrization of perturbations into the action of the theory, expand it up to the second order and then fix the gauge. With further integration out of non-dynamical degrees of freedom one ends up with a quadratic action for 
three
degrees of freedom. This conventional approach has been adopted in numerous works on both cosmological solutions and spherically symmetric settings in different subclasses of DHOST theories.

However, there exists an alternative approach to stability analysis initially suggested for a cosmological setting in Ref.~\cite{DPSV} (we refer to it as a DPSV approach for brevity). The method was put forward within $\mathcal{L}_3$ Horndeski subclass of DHOST theories with the Lagrangian (mostly positive signature; $\kappa = 8\pi G$)
\begin{equation}
\label{eq:L3}
\mathcal{L}_3 = \dfrac{1}{2\kappa}R + F(\pi,X) - K(\pi,X)\square\pi,
\end{equation}
where $\pi$ is the Galileon field, $F$ and $K$ are smooth functions, and
\begin{equation*}
X = - \frac 12 g^{\mu\nu}\partial_\mu\pi\partial_\nu\pi,\quad \square\pi = g^{\mu\nu}\nabla_\mu\nabla_\nu\pi.
\end{equation*}
In $\mathcal{L}_3$ subclass the Galileon field equation (with $F_X = \partial F(\pi,X)/\partial X$, etc.)
\begin{equation}
\begin{aligned}
\label{eq:galeq}
&(F_X - 2 K_{\pi} - K_X \Box{\pi} - 
K_{\pi X} \partial_{\mu}\pi
\partial^{\mu}\pi ) \Box\pi \\
 &- (F_{XX}- 2 K_{\pi X} -  K_{XX} \Box\pi)
\partial^{\mu}\pi\; \nabla_{\mu}\nabla_{\nu}\pi \;\partial^{\nu}\pi \\
&+ K_{X}  \nabla_{\mu}\nabla_{\nu}\pi \; \nabla^{\mu}\nabla^{\nu}\pi
+ K_{XX} \partial^{\mu}\pi\; \nabla_{\mu}\nabla^{\rho}\pi\; \nabla_{\rho}\nabla_{\nu}\pi\; \partial^{\nu}\pi \\
& +
F_{\pi} + (F_{\pi X} - K_{\pi\pi})\partial_{\mu}\pi \partial^{\mu}\pi 
+ K_X R_{\mu\nu} \partial^{\mu}\pi\;\partial^{\nu}\pi= 0,
\end{aligned}
\end{equation}
and Einstein equations
\begin{equation}
\label{eq:eineq}
\begin{aligned}
&G_{\mu\nu}= \kappa K_{X} (\partial_{\mu}\pi\; \nabla_{\nu}\nabla_{\rho}\pi \;\partial^{\rho}\pi + \partial_{\nu}\pi\; \nabla_{\mu}\nabla_{\rho}\pi \;\partial^{\rho}\pi \\
&- \partial_{\mu}\pi \partial_{\nu}\pi \Box\pi 
- g_{\mu\nu} \partial^{\rho}\pi\; \nabla_{\rho}\nabla_{\lambda}\pi \;\partial^{\lambda}\pi) 
 \\
& + \kappa g_{\mu\nu}(F+K_{\pi}\partial_{\rho}\pi \partial^{\rho}\pi)
+ \kappa (F_X - 2 K_{\pi}) \partial_{\mu}\pi \partial_{\nu}\pi, 
\end{aligned}
\end{equation}
involve second derivative terms of both metric and Galileon field (and so do the corresponding linearized equations). It was noted in Ref.~\cite{DPSV} that it is possible to get rid of 
terms with the
second derivatives of metric in the Galileon equation by making use of Einstein equations. So the resulting equation involves the second derivatives of the Galileon field only and upon linearization enables one to reconstruct the derivative part of the quadratic action for the scalar degree of freedom. Originally the first and zeroth derivative terms of both metric and Galileon field were omitted in the resulting Galileon equation, so this approach is suitable for constraining {high momentum and frequency} modes only.
This trick was discussed in detail in Ref.~\cite{gauges1}, where it was proved that DPSV approach in fact corresponds to a specific gauge choice in the quadratic action for perturbations and gives similar result 
{for the stability constraints}
as compared to the conventional approach. The method was further applied and analysed in more general subclasses of Horndeski theories in Refs.~\cite{gauges1,gauges2} for a cosmological background. 

While the conventional approach, where the quadratic action for perturbations is calculated from the scratch, is clear and widely used, in some cases DPSV trick turned out to be a convenient 
alternative~\cite{nogoLMR}. Moreover, DPSV method significantly simplified the stability analysis of the spherically symmetric wormhole solutions in $\mathcal{L}_3$ subclass~\cite{RubakovWorm1,RubakovWorm2,Kolevatov:2016ppi,Evseev:2016ppw}. 
Although DPSV approach was never carefully analysed in a static, spherically symmetric background, its validity in $\mathcal{L}_3$ Horndeski theory is supported by the fact that in $\mathcal{L}_3$ the trick works even for the non-perturbed field equations~\cite{DPSV, gauges1}.

In this note we study the DPSV approach in $\mathcal{L}_3$ Horndeski theory about a static, spherically symmetric background and show that this method amounts to choosing a specific gauge in the quadratic action for perturbations. Hence, we prove the validity of this approach for
stability analysis. 
The latter is a perfect analogy to the cosmological case addressed in Ref.~\cite{gauges1}. We carry out the DPSV trick and find the corresponding gauge choice explicitly in Sec.~\ref{sec:1}. 
Moreover, in Sec.~\ref{sec:2} we check if the DPSV approach works in more general subclasses of Horndeski theories. Here we come across the difference between the cosmological and spherically symmetric settings in Horndeski theory: unlike the case of a homogeneous background, the DPSV trick breaks down as soon as we consider the next $\mathcal{L}_4$ Horndeski subclass. We briefly conclude in Sec.~\ref{sec:3}.

\section{DPSV trick for the spherically symmetric background}
\label{sec:1}

We consider a static, spherically symmetric background with $\pi=\pi(r)$ and
\begin{equation}
\label{eq:backgr_metric}
ds^2 = - A(r)\:dt^2 + \frac{dr^2}{B(r)} + J^2(r) \left(d\theta^2 + \sin^2\theta\: d\varphi^2\right).
\end{equation}
In the linearized field equations
we make use of the standard Regge-Wheeler formalism~\cite{ReggeWheeler} and 
classify perturbations into odd-parity and even-parity types based on their behaviour under two-dimensional reflection. However, we do not consider odd-parity sector since it is trivial and manifestly healthy stability-wise in the $\mathcal{L}_3$ subclass~\eqref{eq:L3}, see e.g. Ref.~\cite{KobaOdd}. As for even-parity sector, upon the expansion in terms of spherical harmonics $Y_{\ell m}(\theta,\varphi)$ the metric perturbations 
$h_{\mu\nu}$ read
\begin{equation}
\label{eq:parametrization_h}
\begin{aligned}
h_{tt}=&A(r)\sum_{\ell, m}H_{0,\ell m}(t,r)Y_{\ell m}(\theta,\varphi), \\
h_{tr}=&\sum_{\ell, m}H_{1,\ell m}(t,r)Y_{\ell m}(\theta,\varphi),\\
h_{rr}=&\frac{1}{B(r)}\sum_{\ell, m}H_{2,\ell m}(t,r)Y_{\ell m}(\theta,\varphi),\\
h_{ta}=&\sum_{\ell, m}\beta_{\ell m}(t,r)\partial_{a}Y_{\ell m}(\theta,\varphi), \\
h_{ra}=&\sum_{\ell, m}\alpha_{\ell m}(t,r)\partial_{a}Y_{\ell m}(\theta,\varphi), \\
h_{ab}=&\sum_{\ell, m} K_{\ell m}(t,r) g_{ab} Y_{\ell m}(\theta,\varphi)\\
&+\sum_{\ell, m} G_{\ell m}(t,r) \nabla_a \nabla_b Y_{\ell m}(\theta,\varphi)\,,
\end{aligned}
\end{equation}
while the perturbed Galileon field reads
\begin{equation}
\label{eq:chi}
\pi (t,r,\theta,\varphi) = \pi(r) + \sum_{\ell, m}\chi_{\ell m}(t,r)Y_{\ell m}(\theta,\varphi),
\end{equation}
where $a,b = \theta,\varphi$, $g_{ab} = \mbox{diag}(1, \;\sin^2\theta)$.
Three
\footnote{
This is the case for $\ell>1$. Both monopole ($\ell=0$) and dipole ($\ell=1$) cases should be considered separately~\cite{KobaEven}.}
of the perturbation variables in eqs.~\eqref{eq:parametrization_h} and~\eqref{eq:chi} can be set to zero by making use of the gauge freedom, as under 
infinitesimal coordinate transformations 
$x^{\mu} \to x^{\mu} + \xi^{\mu}$ with $\xi^{\mu}$ parametrized as
\begin{equation}
\label{eq:xi}
\xi^{\mu} = \Big(T_{\ell m}(t,r), R_{\ell m}(t,r), \Theta_{\ell m}(t,r) g^{ab} \partial_{b}\Big)\,  Y_{\ell m}(\theta,\varphi ),
\end{equation}
perturbation variables change as follows
(in what follows we drop the subscripts $\ell, m$ for brevity)
\begin{equation}
\begin{aligned}
\label{eq:gauge_laws}
H_0 &\rightarrow H_0 + \dfrac{2}{A} \dot{T} - \dfrac{A'}{A} B R, \\
H_1 &\rightarrow H_1 + \dot{R} + T' - \dfrac{A'}{A} T, \\
H_2 &\rightarrow H_2 + 2 B R' + B' R, \\
\beta &\rightarrow \beta + T + \dot{\Theta}, \\
\alpha &\rightarrow \alpha + R + \Theta' - 2 \dfrac{J'}{J} \Theta,\\
K &\rightarrow K + 2 B \dfrac{J'}{J} R, \\
G &\rightarrow G + 2 \Theta, \\
\chi &\rightarrow \chi + B \pi' R,
\end{aligned}
\end{equation}
where prime and dot denote derivatives with respect to $r$ and $t$, respectively. Healthy even-parity sector has two dynamical degrees of freedom (DOF) and we know from e.g. Refs.~\cite{KobaEven,wormhole1} that in $\mathcal{L}_3$ subclass one of these DOFs has a dispersion relation corresponding to that of graviton,
while the other one describes the non-trivial ``scalar'' mode. While the explicit description of even-parity modes in terms of perturbation variables depends on the gauge choice, 
within the DPSV approach we reconstruct the derivative part of the quadratic action for the ``scalar'' mode. 

Let us now carry out the DPSV trick, i.e. eliminate terms with the second derivatives of metric in the Galileon equation~\eqref{eq:galeq} by making use of Einstein equations,
but unlike the original approach we keep all the terms in the resulting Galileon equation. 
So we express $R_{\mu\nu}$ from eq.~\eqref{eq:eineq}, substitute it into $K_X R_{\mu\nu} \partial^{\mu}\pi\; \partial^{\nu}\pi$ in eq.~\eqref{eq:galeq} and then we linearize the resulting Galileon equation.
At this point we do not fix the gauge and derive the resulting linearized Galileon equation in a gauge invariant form:
\begin{equation}
\label{eq:galeq_result}
\begin{aligned}
&q_1 \left(\ddot{\chi} - B\pi' \dot{H}_1\right) + q_{2} \left(\chi'' - \frac{\pi'}2 H_2'\right) \\
& + j^2 q_{3} \left(\chi - B\pi' \left[\alpha  - \frac12 G' + 
\frac{J'}J G\right]\right) \\
& + q_4 H_0' + q_5 K' + q_{6} \chi' + q_7 H_2 + q_{8} \chi 
 = 0,
\end{aligned}
\end{equation}
where $j^2=\ell(\ell+1)$ and it corresponds to the angular part of the Laplacian operator.
We concentrate on the second-derivative terms
in the first and second lines of eq.~\eqref{eq:galeq_result} 
since we focus on constraining
 high energy modes
only. So here we give the explicit form of coefficients $q_1$, $q_2$ and $q_3$ which are relevant for our analysis:
\begin{subequations}
\begin{align}
& q_1 =  A^{-1}\left[ -2 F_X + 4 K_{\pi} + \kappa K_X^2 B^2 {\pi'}^4
\right. \nonumber\\& \left. 
+ 2 K_X (B'\pi' +4 B \frac{J'}{J}\pi' +2 B\pi'' )
+ 2 K_{\pi X} B {\pi'}^2 
\right. \nonumber\\& \left. 
- K_{XX} B {\pi'}^2 (B' \pi' +2 B \pi'')
\right],\\
& q_{2} =  B \left[ 2 F_X - 2 F_{XX} B{\pi'}^2- 4 K_{\pi} 
+ 3 \kappa K_X^2 B^2 {\pi'}^4 
\right. \nonumber\\& \left.
+ (K_{XX} B {\pi'}^2 - 2 K_X ) B \pi' \left(4 \frac{J'}{J} +\frac{A'}{A}\right)
\right. \nonumber\\& \left.
+ 2 K_{\pi X} B {\pi'}^2 
\right],\\
& q_{3} =  J^{-2}\left[- 2 F_X + 4 K_{\pi} + \kappa K_X^2 B^2 {\pi'}^4
\right. \nonumber\\& \left.
+ 2 K_X \left( B\pi' \frac{A'}{A} + B'\pi' + 2 B \frac{J'}{J}\pi' +2 B\pi''\right)  
\right. \nonumber\\& \left.
+ 2 K_{\pi X} B {\pi'}^2 - K_{XX} B {\pi'}^2 (B' {\pi'} + 2 B\pi'')
\right].
\end{align}
\end{subequations}
Let us note that within the original DPSV approach one would 
drop $\dot{H}_1$, $H_2'$, $j^2\alpha$ and $j^2 G'$ terms in eq.~\eqref{eq:galeq_result} since they appear to be the first-derivative quantities from
the covariant point of view, see eq.~\eqref{eq:parametrization_h}:
\begin{equation}
\dot{H}_1 \sim \partial_t h_{tr}, \;\; 
H_2' \sim \partial_r h_{rr}, \;\;
j^2 \alpha \sim \partial_a h_{ra}, \;\;
j^2 G' \sim \partial_r h_{ab}.
\end{equation}
However,
careful analysis of the constraint equations shows that 
$H_1$ and $H_2$ perturbation variables 
behave as the first-derivative objects.
Indeed,
the constraint equations obtained from a full-fledged analysis of the quadratic action for perturbations, carried out e.g. in Refs.~\cite{KobaEven,wormhole1}, read:
\begin{subequations}
\label{eq:constraints}
\begin{align}
\label{eq:H1constraint}
 j^2 H_1 =& j^2 {\dot \alpha} + 2\kappa\Xi {\dot {\chi}}' 
 - \Omega {\dot H_2} +2J^2 \dot{K}' +\dots,\\
\label{eq:H2constraint}
H_2' =&  \frac1{\Omega}\left(2\kappa\Xi \chi'' + 2 J^2 K''+ 2 j^2  \alpha' + j^2 B^{-1} H_2 
+\dots\right),
\end{align}
\end{subequations}  
where dots denote terms with the first and zeroth order derivatives and
\begin{equation}
\Omega = 2 JJ' + \kappa \Xi\pi', \qquad \Xi = K_X BJ^2 {\pi'}^2.
\end{equation}
We see that both $H_1$ and $H_2$ are of the same order as derivatives of $\alpha$ and $\chi$. Therefore, extra terms $B\pi' \dot{H}_1$ and $\frac{\pi'}2 H_2'$ in eq.~\eqref{eq:galeq_result} have to be treated as second-derivative objects and they may contribute
to the dispersion relation for $\chi$
(cf. Ref.~\cite{gauges1}). 
Another way to see that $H_1$ and $H_2$ behave as the first-derivative quantities is to notice that they both involve first derivatives of gauge parameters in the transformation laws~\eqref{eq:gauge_laws}.
At this point it is not entirely obvious why does the DPSV trick give the correst result at all, since it dictates to omit seemingly relevant terms in the linearized equations.
Let us now show that there is a gauge choice which makes the extra terms
with $\dot{H}_1$, $H_2'$, $j^2\alpha$ and $j^2 G'$ simultaneously
vanish in eq.~\eqref{eq:galeq_result}.

First, we partially fix the gauge in eq.~\eqref{eq:galeq_result} by making use of functions $T(t,r)$ and $\Theta(t,r)$ and set
$\beta=G=0$,
while the invariance under transformations with 
gauge function $R(t,r)$ is still intact. The constraints~\eqref{eq:constraints} are also exactly $R$-invariant, and the terms given in both eq.~\eqref{eq:H1constraint} and~\eqref{eq:H2constraint} yield
the second and first 
derivatives of $R(t,r)$ under gauge transformations~\eqref{eq:gauge_laws}.
What is more, both radial and angular parts of constraints are $R$-invariant independently. Hence, 
to the leading order in derivatives of $R$ the angular parts of constraints~\eqref{eq:constraints} give the following relations, respectively:
\begin{equation}
\label{eq:H1H2relation}
H_1 = \dot{\alpha}, \qquad H_2 = 2B \alpha'.
\end{equation}
These relations enable us to 
rewrite eq.~\eqref{eq:galeq_result} as follows:
\begin{multline}
\label{eq:galeq_finL3}
q_1 (\ddot{\chi} - B\pi' \ddot{\alpha}) + q_{2} \left(\chi'' - B\pi'\alpha''\right) \\
 +  j^2 q_{3} \left(\chi - B\pi' \alpha \right) +\dots
 = 0.
\end{multline}
Therefore, 
as soon as we fix the residual gauge freedom by making $\alpha =0$ with a proper choice of $R(t,r)$ function, the extra terms 
$B\pi' \dot{H}_1$ and $\frac{\pi'}2 H_2' $ in eq.~\eqref{eq:galeq_result} vanish automatically due to relations~\eqref{eq:H1H2relation} and we are back to the result given by the original DPSV trick. 
Let us note that the dispersion relation for $\chi$, which follows from eq.~\eqref{eq:galeq_finL3}, coincides with that one 
obtained within a conventional approach in e.g. Ref.~\cite{KobaEven}. 

Thus, we have explicitly shown that DPSV trick in a static, spherically symmetric background corresponds to an implicit gauge choice
$$
\alpha=\beta=G = 0.
$$

\section{DPSV trick in $\mathcal{L}_4$}
\label{sec:2}

Let us consider the DPSV trick in a broader $\mathcal{L}_3+\mathcal{L}_4$ subclass of Horndeski theory:
\begin{equation}
\label{eq:L3L4}
\begin{aligned}
\mathcal{L}_3+\mathcal{L}_4 &= F(\pi,X) - K(\pi,X)\square\pi 
+ G_4(\pi,X)R \\
&+ G_{4X}(\pi,X) \left[\left(\Box\pi\right)^2-\pi_{;\mu\nu}\pi^{;\mu\nu}\right].
\end{aligned}
\end{equation}
The case considered in the previous section corresponds to 
$G_{4}(\pi) =1/(2\kappa)$. The non-trivial function $G_{4}(\pi,X)$
yields 
additional terms with second derivatives of metric in the Galileon field equation:
\begin{equation}
\begin{aligned}
\label{eq:galileonL3L4}
&(G_{4\pi} + G_{4X} \Box\pi + G_{4\pi X}\partial_{\mu}\pi\partial^{\mu}\pi - G_{4XX} \partial^{\mu}\pi\nabla_{\mu}\nabla_{\nu}\pi \partial^{\nu}\pi ) \; R \\
& + (K_X - 4 G_{4\pi X} - 2G_{4XX} \Box\pi) R_{\mu\nu}\partial^{\mu}\pi\partial^{\nu}\pi \\
& + 4 G_{4XX} R_{\mu\nu} \partial_{\mu}\pi \nabla_{\rho}\nabla^{\nu}\pi \partial^{\rho}\pi - 2 G_{4X} R_{\mu\nu} \nabla^{\mu}\nabla^{\nu}\pi \\
& + 2 G_{4XX} R_{\mu\nu\lambda\rho} \partial^{\mu}\pi \partial^{\lambda}\pi \nabla^{\nu}\nabla^{\rho}\pi + \dots = 0,
\end{aligned}
\end{equation}
where dots stand for terms without second derivatives of metric.
The corresponding Einstein equations in $\mathcal{L}_4$ read as follows:
\begin{equation}
\begin{aligned}
\label{eq:einsteinL3L4}
& -2 G_{\mu\nu} G_4 + G_{4X} (R \partial_{\mu}\pi \partial_{\nu}\pi 
- 2 R_{\mu\rho} \partial_{\nu}\pi \partial^{\rho}\pi) \\
&- 2 R_{\nu\rho} \partial_{\mu}\pi \partial^{\rho}\pi
+ 2 g_{\mu\nu} R_{\rho\lambda} \partial^{\rho}\pi \partial^{\lambda}\pi
- 2 R_{\mu\lambda\nu\rho} \partial^{\lambda}\pi \partial^{\rho}\pi)\\
&+ \dots = 0,
\end{aligned}
\end{equation}
where we again drop the
terms without the second derivatives of metric.

The DPSV trick in $\mathcal{L}_3+\mathcal{L}_4$ Horndeski theory
has been already discussed in Ref.~\cite{gauges1}. It was proved that the trick does not go through in $\mathcal{L}_3+\mathcal{L}_4$ in the case of the general background: the terms with the second derivatives of metric in the linearized Galileon equation~\eqref{eq:galileonL3L4} are not a linear combination of the corresponding second derivative terms in the linearized Einstein 
equations~\eqref{eq:einsteinL3L4}. However, it turned out that once one considers the 
cosmological background with a homogeneous Galileon field and spatially flat FLRW metric, the trick works again and gives the correct result for the derivative part of the action for perturbations. In this section we aim to find out if the DPSV approach also works at the linearized level of $\mathcal{L}_3+\mathcal{L}_4$ in the case of static, spherically symmetric background. 

We return to
a static, spherically symmetric background with Galileon field
$\pi=\pi(r)$ and metric~\eqref{eq:backgr_metric}. In full analogy with the cosmological case the number of non-trivial components of $R_{\mu\nu\lambda\rho}$ in the Galileon field equation upon linearization get reduced as soon as we specify the background setting 
($0,1,2,3=t,r,\theta,\varphi$):
\begin{equation}
\label{eq:gallinearizedL3L4}
\begin{aligned}
&Q_1 R^{(1)}_{0101} + Q_2 \left(R^{(1)}_{0202} + \sin^{-2}\theta\: R^{(1)}_{0303}\right) + Q_3 R^{(1)}_{2323} \\
&+ Q_4 \left(R^{(1)}_{1212} + \sin^{-2}\theta\: R^{(1)}_{1313}\right) + \dots = 0,
\end{aligned}
\end{equation}
where $Q_i$ are independent coefficients in terms of function $G_4$
and its derivatives, whose explicit form is irrelevant at this point, and dots denote terms without second derivatives of metric perturbations.
As before we aim to eliminate the second derivatives of metric perturbations in the Galileon field equation by making use of Einstein equations, so that there are second derivative terms of Galileon perturbation $\chi$ only. From eq.~\eqref{eq:gallinearizedL3L4} we see that in order to achieve this one has to 
substitute
the following four combinations:
$R^{(1)}_{0101}$, $\left(R^{(1)}_{0202} + \sin^{-2}\theta\: R^{(1)}_{0303}\right)$, $R^{(1)}_{2323}$ and 
$\left(R^{(1)}_{1212} + \sin^{-2}\theta\: R^{(1)}_{1313}\right)$.
The same combinations appear in the linearized Einstein equations upon 
contracting them with
\begin{equation}
\begin{aligned}
\label{eq:background_comb}
&g^{\mu\nu}, \quad \partial^{\mu}\pi \partial^{\nu}\pi, \quad
\nabla^{\mu}\nabla^{\nu}\pi, \\
&\partial^{\mu}\pi \:\nabla^{\nu}\nabla^{\rho}\pi \:\partial_{\rho}\pi, \quad 
\nabla^{\mu}\nabla^{\rho}\pi \;\nabla_{\rho} \nabla^{\nu}\pi, 
\quad \mbox{etc.}
\end{aligned}
\end{equation}
Let us note that contraction of Einstein equations with 
any of the objects in~\eqref{eq:background_comb} in fact
gives different linear combinations of Einstein equations' components.
For example, contraction with $g_{\mu\nu}$ gives the sum of all linearized Einstein equations, while contraction with $\partial^{\mu}\pi \partial^{\nu}\pi$ results in obtaining only $11$-component. 
The key observation we come across for the spherically symmetric background~\eqref{eq:backgr_metric} is that contractions with background objects 
from~\eqref{eq:background_comb} give
only three independent combinations of linearized Einstein equations' components. The latter makes it impossible to express four combinations of linearized Riemann tensor
in eq.~\eqref{eq:gallinearizedL3L4} from the linearized Einstein equations. 
So the DPSV does not work in
$\mathcal{L}_4$ Horndeski theory in a static, spherically symmetric setting.

The fact that there are only three independent combinations of the linearized Einstein equations in the spherically symmetric case 
appears natural since metric~\eqref{eq:backgr_metric}
is $S^2$-symmetric and, hence, $22$-component and $33$-component of linearized Einstein equations always appear in a specific linear combination and cannot be separated. 
Similar reasoning applies to the cosmological setting: in that case flat FLRW metric has $R^3$-symmetry, so there are two independent combinations of the linearized Einstein equations, i.e. $00$-component and e.g. trace, since all three spatial components appear 
in one linear combination. 
However, due to the symmetry of the space-time there appear only two combinations of 
$R_{\mu\nu\lambda\rho}$ components in the linearized Galileon equation, so in that case there are enough independent combinations of linearized Einstein equations to solve the system for two variables.

\section{Conclusion}
\label{sec:3}
In this paper we have analysed the alternative DPSV approach to deriving 
stability conditions for high momenta modes about
static, spherically symmetric background in Horndeski theories. 
We showed that the DPSV trick amounts to a specific gauge choice in
the Lagrangian for perturbations and, hence, proved its validity in $\mathcal{L}_3$ subclass.
In this perspective the case of spherically symmetric setting 
is highly
analogous to the cosmological FLRW background.

However, in contrast to the cosmological case, it was shown that the DPSV approach breaks down in a
spherically symmetric background as soon as we extend to 
$\mathcal{L}_3+\mathcal{L}_4$ subclass of Horndeski theory. 
The latter result appears natural
since in the $\mathcal{L}_4$ subclass even parity modes do not decouple 
and both
``graviton'' and ``scalar'' modes  have non-trivial dispersion relations~\cite{KobaEven,wormhole1},
so it is not quite clear 
how the DPSV approach could possibly detach these two modes.

\section*{Acknowledgements}
This work has been
supported by Russian Science Foundation grant
19-12-00393.



\end{document}